\documentclass[%
  ,reprint
  ,letterpaper
  ,superscriptaddress
  ,showkeys
  ,amsmath,amssymb
  ,aps
  ,nofootinbib
]{revtex4-1}

\usepackage[pdftex]{graphicx}
\usepackage{dcolumn}
\usepackage{bm}

\begin{document}

\title{Anomalous popularity growth in social tagging ecosystems}

\author{Yasuhiro Hashimoto}
\email{hashi@sacral.c.u-tokyo.ac.jp}
\affiliation{University of Tokyo, Graduate School of Arts and Sciences, Tokyo, 153-8902, Japan}

\author{Mizuki Oka}
\affiliation{University of Tsukuba, Faculty of Engineering, Information and Systems, Tsukuba, 305-8573, Japan}

\author{Takashi Ikegami}
\affiliation{University of Tokyo, Graduate School of Arts and Sciences, Tokyo, 153-8902, Japan}

\keywords{Social tagging, Yule--Simon process, Preferential attachment, Popularity growth, Large deviation}

\begin{abstract}
  In social tagging systems, the diversity of tag vocabulary and the popularity of such tags continue to increase as they are exposed to selection pressure derived from our cognitive nature and cultural preferences. This is analogous to living ecosystems, where mutation and selection play a dominant role. Such population dynamism, which yields a scaling law, is mathematically modeled by a simple stochastic process---the Yule--Simon process, which describes how new words are introduced to the system and then grow. However, in actual web services, we have observed that a large fluctuation emerges in the popularity growth of individual tags that cannot be explained by the ordinary selection mechanism. We introduce a scaling factor to quantify the degree of the deviation in the popularity growth from the mean-field solution of the Yule--Simon process, and we discuss possible triggers of such anomalous popularity behavior.
\end{abstract}

\flushbottom
\maketitle

\section{introduction}
{\itshape Social tagging systems}~\cite{Pink2005} are a kind of information retrieval system that has been widely adopted in modern web services wherein individuals share information resources such as photos, movies, articles, web bookmarks, and so on. The system allows people to associate their resources with a set of arbitrary short texts, called {\itshape tags}, as annotation for future retrieval. What kind of words may be used as a tag is totally up to service users themselves; consequently, the vocabulary of tags continues growing to reflect the diversity of our activities of daily living, including attentions, awareness, preferences, and creations, as if we have coded them democratically into computable entities. From such a point of view, we naturally come to the following two questions:
\begin{enumerate}
\item When and at what rate do we create a new word?
\item How frequently do we use each word temporally and cumulatively?
\end{enumerate}
The first question implies a sense of the expansion of our cognitive space through the creation of a novel idea, and the second question implies a sense of the form of the universe in which each of us dwells. These questions may remind us of one mathematical model---the Simon process~\cite{Simon1955}, which was introduced by Simon in 1955 to explain interdisciplinary popularity dynamics such as word-use frequencies, city sizes, scientific outcomes, and so on. The process yields a sequence of symbols by adding one at a time; here we consider each distinct symbol as a tag. In the Simon process, we may assume two rules for a tag that is newly added to the sequence as follows:
\begin{enumerate}
\item With probabililty \(\alpha\), which we call {\it novelty rate}, the tag is a brand new word that has not yet appeared in the sequence.
\item With complementary probability \(1-\alpha\), the tag is selected among the existing tags in the sequence according to a certain preference.
\end{enumerate}
It is obvious that these two assumptions offer direct answers to each question raised above, that is, how new words are introduced to the system and how they grow---increase their popularity. Our interest here is particularly in the latter question---the growth of tags' popularity, where we have found that the actual web services exhibit a notably large diversity in the growth patterns of tags. Several existing studies~\cite{Golder2006,Cattuto2006,Halpin2007,Cattuto2007,Cattuto2009} have reported social tagging analyses from a variety of statistical approaches, however, diversity in individual tag growth has not been explicitly discussed before, although it is a non-negligible nature of tags, reflecting the selection pressure exerted by human beings, based on our cognitive nature and cultural preferences of the time. We expect that revealing the growth dynamics of tags will lead to insights on our novelty adoption, attention shift, and emergence of new lifestyle niche~\cite{Kauffman2000,Tria2014}.

This study shows that a well-defined mathematical model, the Yule--Simon process---the specialization of the Simon process mentioned above---can reproduce to a certain extent social tagging behavior in real web services; however, it has a limitation in explaining anomalous growth patterns of tags' popularity. In order to clarify the limitation, we formalize an analytical form of the popularity growth in the Yule--Simon process that innately contains fluctuations and introduce a scaling factor to quantify the deviation from the mean-field solution. This scaling factor will provide perspective into the growth process of tags. In the second section, we give a brief but essential explanation of the Simon and Yule--Simon processes to establish a base for the following discussion. In the third section, we give an empirical analysis of real web services. In the fourth section, we discuss the discrepancy between the data and the theory, and we provide a conclusion in the last section.

\section{The Simon and Yule--Simon processes}
\label{sec:YS}
The Simon process is based on a review of the Yule process~\cite{Yule1925} that was introduced by Yule in 1925 to explain the power-law population distribution observed in biological species over genera~\cite{Willis1922}. While the models differ slightly in detail, they share a common idea in increasing diversity and population of system components. Here we elaborate the difference between the two models and clarify the definition of the model we used here---the Yule--Simon process.

In the Simon process, a tag is added to the tag sequence at every time step $t$, and the length of the sequence $N(t)$ grows linearly; $N(t)=t$ by definition. At every time step, we find a new tag that has never appeared in the sequence with probability $\alpha$, or that already exists in the sequence with the complementary probability $1-\alpha$, or $\bar{\alpha}$. In a realistic situation, this novelty rate $\alpha$ may depend on time; in such a case, we use $\alpha(t)$. Hence, choosing $\alpha$ makes the vocabulary size increase as $\sum_t\alpha(t)$. In contrast, the latter choice, $\bar{\alpha}$, increases the existing tags' popularity, that is, the cumulative number of occurrences. In the latter case, Simon defined the probability of an arbitrary tag belonging to {\it class-$n$}, which is a set of distinct tags that appear exactly $n$ times in the sequence, as proportional to the number of total occurrences of class-$n$ tags. That is, the rate equation of $k_n(t)$, the vocabulary size of class-$n$ at $t$, is described as follows:
\begin{align}
  k_1(t+1)&=k_1(t)+\alpha(t)-\bar{\alpha}(t)\mathcal{P}_1(t),\nonumber\\
  k_n(t+1)&=k_n(t)+\bar{\alpha}(t)[\mathcal{P}_{n-1}(t)-\mathcal{P}_n(t)]\quad (n\ge 2),
  \label{eq:rate_equation_for_tag_class}
\end{align}
where $\mathcal{P}_n(t)$ is the probability that class-$n$ is chosen at right after $t$, defined as
\begin{align}
  \mathcal{P}_n(t)\equiv\frac{nk_n(t)}{N(t)}.
  \label{eq:probability_of_class_selection}
\end{align}
These equations result in a power-law popularity distribution known as Zipf's law~\cite{Zipf1935} in conventional conditions. For example, if $\alpha$ has a constant value, the exponent of Zipf's law becomes $\bar{\alpha}$; meanwhile, if $\alpha$ decays temporally as $\alpha(t)\propto t^{\beta-1}$, where $\beta$ is a positive constant below one, the exponent of Zipf's law becomes $\beta^{-1}$~\cite{Simon1955}.

As we see, Simon postulated how the class should be selected but not how an individual tag should be selected. Therefore, the Simon process is not well defined for modeling the growth of individual tags. Yule's postulate---every biological genus having the same number of species grows at the same rate---is interpreted in Simon's framework as every tag in the same class having the same probability of selection. This additional assumption complements a requirement for describing the growth of individual tags. Now we define the Simon process with Yule's growth rate as the Yule--Simon process, where the probability of picking tag $i$ out of class-$n$ at right after $t$ is defined as
\begin{align}
  P_i(t)\equiv\left.\frac{\mathcal{P}_n(t)}{k_n(t)}\right|_{n=n_i(t)}=\frac{n_i(t)}{N(t)},
  \label{eq:probability_of_tag_selection}
\end{align}
where $n_i(t)$ is the popularity of tag $i$ at $t$. This selection probability proportional to its popularity is well-known as {\it preferential attachment} dynamics. Consequently, the growth of each tag is described not using the rate equations but as follows:
\begin{align}
  n_i(t+1)&=n_i(t)+\bar{\alpha}(t)P_i(t).
  \label{eq:growth}
\end{align}
In typical cases, $\alpha(t)$ is time-decaying and its value is very small in mature web services, $\sim 10^{-2}$ or below, as we will see in the next section. Assuming a continuum limit of $t$ and $n_i(t)$ and replacing $\bar{\alpha}(t)$ with one, we obtain a solution
\begin{align}
  n_i(t)=\left(\frac{t}{t_i}\right)^{\bar{\alpha}}\sim\frac{t}{t_i},
  \label{eq:growth_solution}
\end{align}
where $t_i$ is the time that tag $i$ was created, that is, used for the first time in the service, and $t\ge t_i$ and $n_i(t_i)=1$ by definition.

\begin{figure}[t]
\centering
\includegraphics[width=.9\linewidth]{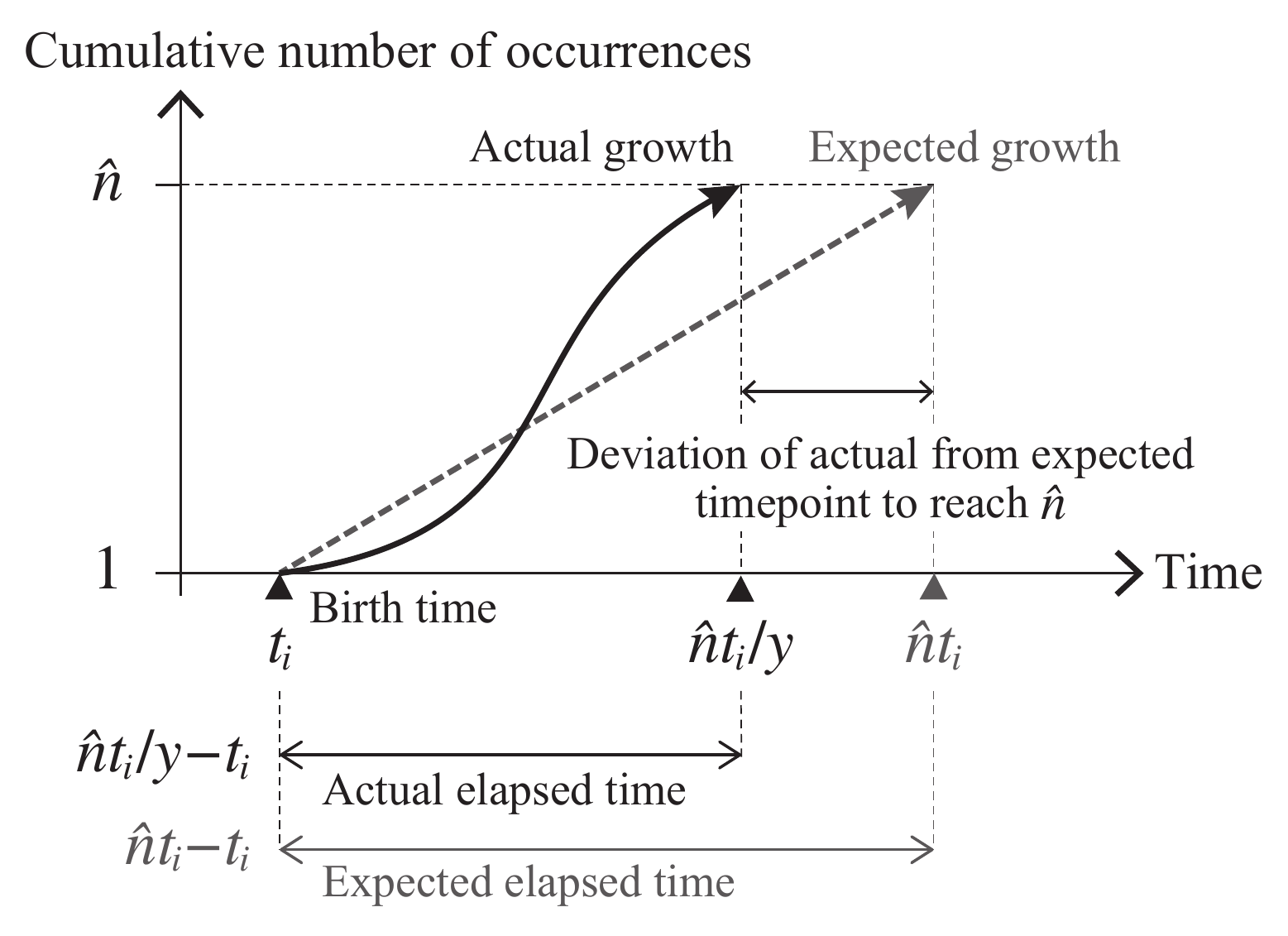}
\caption{Schematic diagram of the time deviation from the mean-field prediction in tag growth. The actual growth curve likely has a certain degree of fluctuation around the expected growth curve. This example shows ``faster'' growth than the prediction for given $\hat{n}$.}
\label{fig:diagram}
\end{figure}

In the actual discrete stochastic process, we observe a certain degree of fluctuation around the mean-field solution (\ref{eq:growth_solution}). Such fluctuation can be formalized in two different aspects---size and time---as follows: We define the size deviation scaling factor, say $x$, as the amount by which the actual observed popularity at a specific timepoint is $x$-times larger than the mean-field solution. Then, the probability density distribution of $x$ for the Yule--Simon process is approximately derived as follows~\cite{Hashimoto2016}:
\begin{align}
  p(x)&\sim\left(1-\frac{1}{\hat{n}}\right)^{x\hat{n}-1},
  \label{eq:PD_to_be_xn_at_t_}
\end{align}
where $\hat{n}$ is a target value of the popularity. Using a small value of $\hat{n}$, we will measure the deviation in the early stage of tag popularity growth. In contrast, using a large value of $\hat{n}$ allows the evaluation of the deviation in long-term growth. The detail of the derivation is shown in Appendix (\ref{sec:px}). In a similar way, we also define the timepoint deviation scaling factor, say $y$, as that the actual observed timepoint where the tag reaches $\hat{n}$ $y$-times faster than the mean-field solution (Fig.~{\ref{fig:diagram}}). The timepoint deviation scaling factor can be easily converted into the elapsed-time deviation scaling factor, say $y'$. We will evaluate the temporal deviation using $y'$ instead of $y$. The probability density distribution of $y'$ for the Yule--Simon process is approximately derived as follows:
\begin{align}
  q(y')&= yp(y')\nonumber\\
  &=\frac{\hat{n}y'}{y'+\hat{n}-1}\left(\frac{\hat{n}-1}{y'+\hat{n}-1}\right)^{\hat{n}},
  \label{eq:PD_to_be_y2t_with_correction_}
\end{align}
where $y$ multiplied by $p(y')$ in the first line is a correction term for the observation bias in the finite data. Details of the derivation are also shown in Appendix (\ref{sec:py}).

\begin{figure*}[ht]
\centering
\includegraphics[width=.95\linewidth]{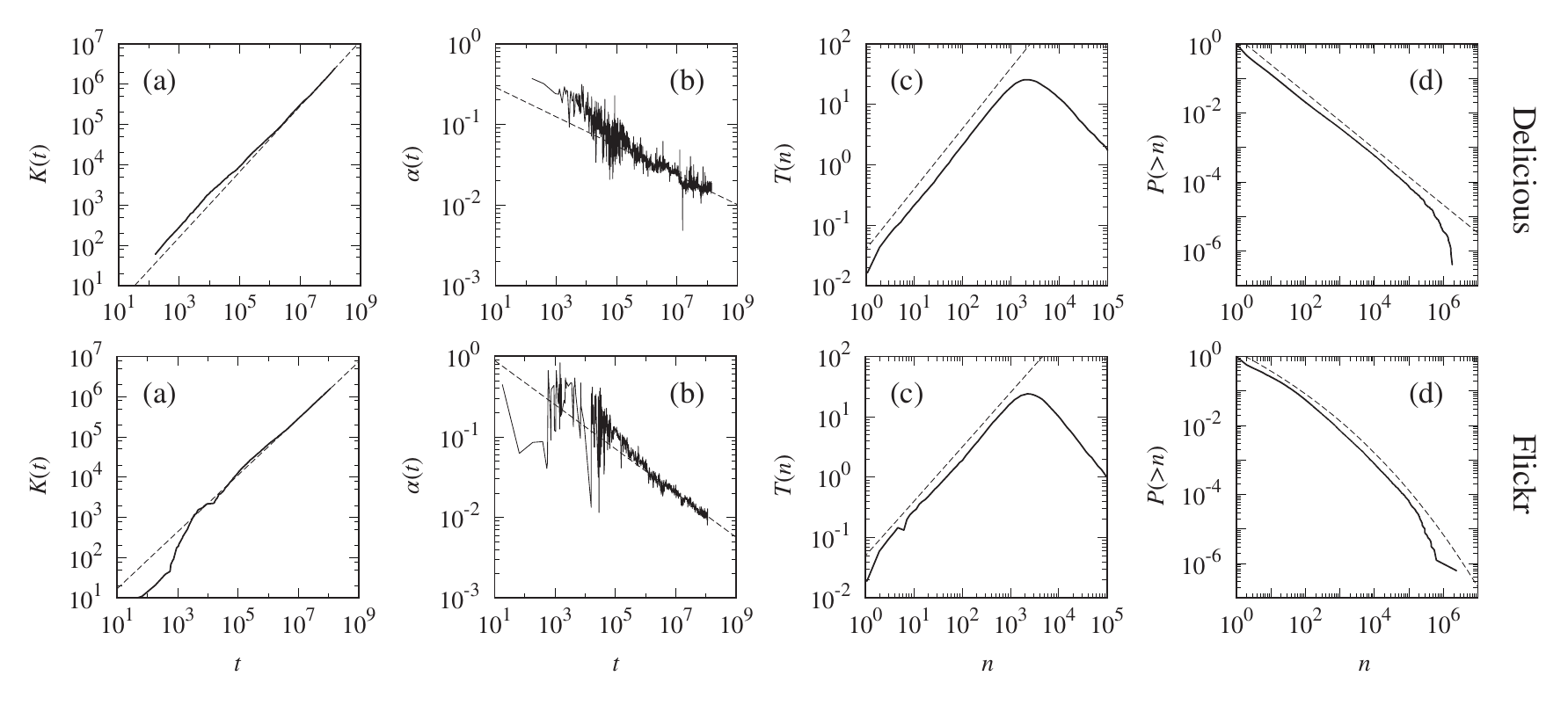}
\caption{Statistics on the vocabulary growth and the class selection in the real data. Top and bottom sets of figures show the results from Delicious and Flickr, respectively. In (a)---the growth of the vocabulary size and (b)---the transition of the novelty rate, each data point corresponds to a daily result, where the horizontal value is the cumulative number of annotations at the end of the day. Here, $K(t)$ and $\alpha(t)$ are the vocabulary size at the end of the day and the daily average novelty rate, respectively, and every successive daily result is connected by solid lines. Dashed lines are a power curve fitted for $t>10^6$ by using the ordinary least-squares method; fitted values are $\beta=0.813\pm 0.126\%$ and $\gamma=0.182\pm 2.12\%$ for Delicious and $\beta=0.704\pm 0.0536\%$ and $\gamma=0.274\pm 1.47\%$ for Flickr. (c) The weighted frequency of class selections $T(n)$ against the class index $n$. Dashed lines, shown to guide the eye, have a slope of 1 for Delicious and 0.9 for Flickr in a double logarithmic scale. (d) Complimentary cumulative probability of the class-$n$ vocabulary size. Dashed lines are shown to guide the eye, following a power curve $P(>n)\propto n^{-0.813}$ in Delicious and a stretched exponential curve $P(>n)\propto \exp(-7.8n^{0.070})$ in Flickr.}
\label{fig:heaps_and_zipf_law}
\end{figure*}

\section{Empirical analysis}

\subsection{Datasets}
We now examine two huge datasets gathered from actual web services---Delicious and Flickr---by G\"{o}rlitz and others~\cite{Gorlitz2008}. The first service is provided for sharing web bookmarks that can be tagged by multiple users. The second service is provided for sharing photos that are tagged only by the user who posted the photo. These two types of tagging are distinguished, respectively, as broad-tagging and narrow-tagging systems~\cite{Vanderwal2005}; however, this distinction is not taken into account in this study. The datasets consist of a list of annotations, where each annotation includes a time stamp of when the annotation was created, the user name who annotated the resource, the resource index, and the string of the used tag. In most cases, multiple annotations are made simultaneously for a single resource. Here, physical time, user names, resource indices, and simultaneity of multiple annotations are irrelevant to our analysis and we focus only on a temporally sorted sequence of tag strings wherein the cumulative number of annotations is regarded as pseudo time---one annotation corresponds to one time step. The total numbers of distinct tags $K$ and annotations $N$ included in the datasets are $K\sim 2.5\times 10^6$ and $N\sim 1.4\times 10^8$ for Delicious and $K\sim 1.6\times 10^6$ and $N\sim 1.1\times 10^8$ for Flickr, respectively.

\subsection{Heaps' law and Zipf's law}
First, we briefly demonstrate that the well-known stylized facts observed in human linguistic activities---Heaps' law~\cite{Heaps1978} and Zipf's law---are also observed in real social tagging systems to a limited degree, and they are reasonably explained within the framework of the Yule--Simon process. Heaps' law is a sub-linear relationship between the total number of annotations (pseudo time) and the vocabulary size as follows:
\begin{align}
  K(t)=K_0 t^{\beta},
\end{align}
where $K(t)$ is the vocabulary size at $t$, $K_0$ is the initial vocabulary size, and $\beta$ is a positive constant below one. If the novelty rate is time-decaying following the power-law relationship as $\alpha(t)\propto t^{-\gamma}$, where $\gamma$ is a positive constant below one, the resulting vocabulary growth should follow Heaps' law with $\beta=1-\gamma$ from the definition of $\alpha(t)=dK/dt$. Hence, this statistical law of vocabulary growth can be naturally incorporated into the Yule--Simon process as a boundary condition in novelty production, whose mechanism we will not discuss in this study. Figure \ref{fig:heaps_and_zipf_law} (a) and (b) show, respectively, transitions of $K(t)$ and $\alpha(t)$ in the data. We can see that the empirical vocabulary growth in both services follows Heaps' law, and their time-decaying novelty rates come down to the order of $10^{-2}$ as the service matures; furthermore, it seems to get even lower into the future. So, our assumption of sufficient small $\alpha$ in defining the deviation scaling factors would be valid in the large limit of $t$.

Zipf's law is a direct result from Eqs.~(\ref{eq:rate_equation_for_tag_class}) and (\ref{eq:probability_of_class_selection}). Therefore, by counting selections of class-$n$ in the evolving process, we can verify whether proportional class selection to $n$ has actually occurred~\cite{Newman2001}. Figure \ref{fig:heaps_and_zipf_law} (c) and (d) show, respectively, the weighted frequency of class selections against $n$ and the resulting class size distributions. The dashed straight lines drawn in (c) have slopes $\psi=1$ (Delicious) and $\psi=0.9$ (Flickr) in a double logarithmic scale, which suggests that linear and non-linear preferential attachment is working in each case, respectively. As a result, Delicious shows a power-law class size distribution; meanwhile, Flickr follows a stretched exponential distribution~\cite{Krapivsky2001}.

Looking into the parameters of both distributions in detail, the power exponent in Delicious is roughly equal to Heaps' exponent $\beta$, which is in accordance with Simon's consequence mentioned in \ref{sec:YS}. On the other hand, the slope of the non-linear preferential attachment is known to appear in the shape parameter of the stretched exponential distribution as $\exp(-Cx^{1-\psi})$ when $1/2<\psi<1$, wherein the least-squares fitting value of $\psi$ roughly matches the slope in (c) for Flickr. Zipf's law has not held in this case, but the result still seems consistent and reasonable. In summary, when looking at the popularity growth of tags from the usual statistical aspects, we find no stunning facts that go beyond the preferential attachment framework.

\subsection{Large deviation in popularity growth}
\begin{figure*}[t]
\centering
\includegraphics[width=.8\linewidth]{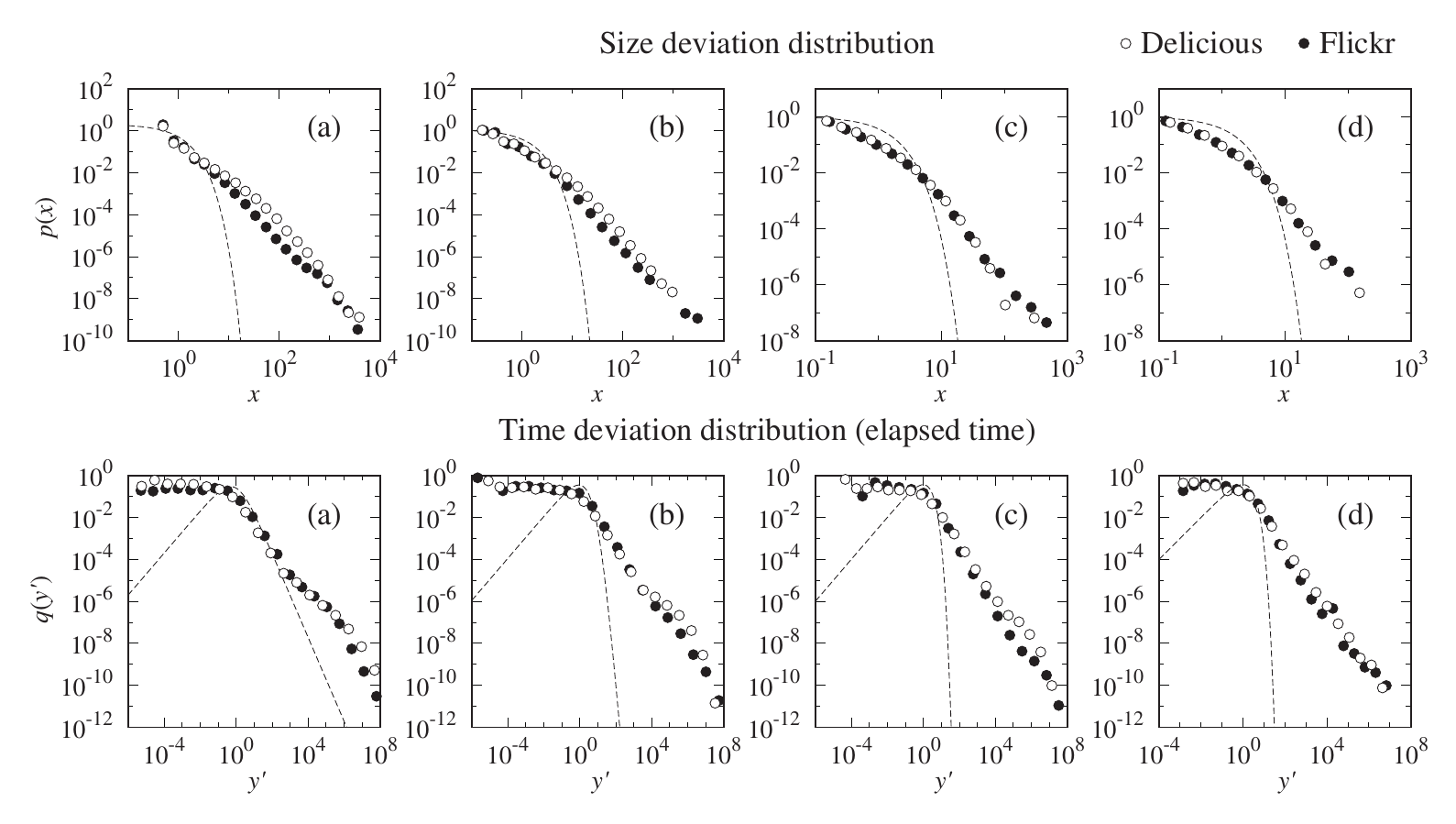}
\caption{Probability density distribution of the size and time deviation scaling factors in the real data. (a) to (d) correspond to the cases of $\hat{n}=2$, 10, $10^2$, and $10^3$, respectively. White and black dots show the results from Delicious and Flickr, respectively, where logarithmic binning was used with 20 bins between the max and min values. Dashed lines are the theoretical curves (\ref{eq:PD_to_be_xn_at_t_}) and (\ref{eq:PD_to_be_y2t_with_correction_}) for given $\hat{n}$.}
\label{fig:pdf.empirical}
\end{figure*}

Now, let us investigate the popularity growth of individual tags. Figure \ref{fig:pdf.empirical} shows the comparison between empirical results from the two datasets (white and black dots) and the theoretical curves (\ref{eq:PD_to_be_xn_at_t_}) and (\ref{eq:PD_to_be_y2t_with_correction_}) (dashed lines) with varying $\hat{n}$ as 2, 10, $10^2$, and $10^3$. Overall, the shapes of the empirical results are significantly skewed compared to the theory in every case, and there exist a considerable amount of tags that grow extremely rapidly or slowly, far beyond the innate fluctuation of the Yule--Simon process. In addition, both datasets have interestingly similar distribution shapes; nevertheless the natures of the services and their architecture are fairly different. Let us recall the sense of the notion $x$ for example; if we observe tag $i$ having a large value of $x$, say $x=10^3$, it means that the tag gained $10^3\hat{n}$ occurrences by time step $\hat{n}t_i$, at which only $\hat{n}$ occurrences are expected if we assume the ordinary preferential attachment. Such a gain of popularity is extraordinarily rare in the Yule--Simon process, and we need another mathematical or phenomenological treatment for those tags. We address this question with two more analyses in the following.

\begin{figure*}[t]
\centering
\includegraphics[width=1\linewidth]{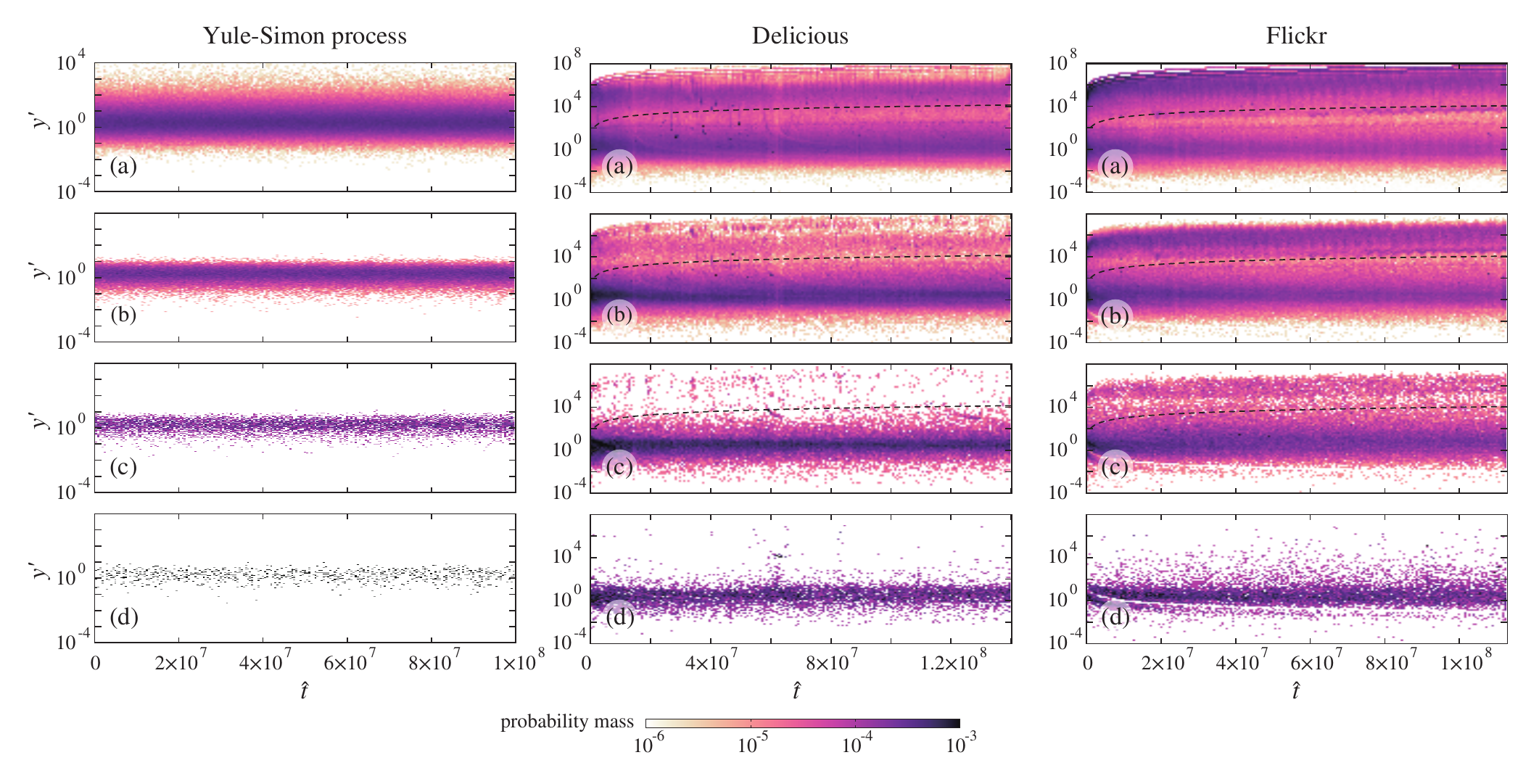}
\caption{Relationship between $y'$ of qualified tags and the timepoint at which they reached given $\hat{n}$: (a) to (d) correspond to the cases of $\hat{n}=2,10,10^2,10^3$, respectively. The color hue indicates the probability mass of tags counted in the heat map cell, normalized by the total number of qualified tags. The cell size is log-binned with 80 bins vertically and linear-binned with 200 bins horizontally between the max and min values. Black dashed lines are $y'=\hat{t}/10^4$ for eye guide.}
\label{fig:heatmap.t_vs_y}
\end{figure*}

Figure \ref{fig:heatmap.t_vs_y} shows the time-dependency of $y'$; the vertical axis is $y'$ of each qualified tag---a tag that gained given $\hat{n}$ within the data period, and the horizontal axis is the timepoint at which the tag reached $\hat{n}$. The color hue of the heat map cells indicates the number of tags that fall within the cell, normalized by the total number of qualified tags. The cell size is log-binned vertically and linear-binned horizontally, so the color hue displays not a probability density but a probability mass in the cell. Looking at the mass distribution, we can capture the absolute portion of tags extending across a wide range of $y'$. We find here again extremely rapidly growing tags in the real data, and moreover, they have formed a sort of temporal structure like a stratum. For example, see the area between $y'=10^4$ and $10^8$ in (a) of the empirical results. We identify in that area a dense layer apart from the other dense layer around $y'=1$, which would be a group of relatively ordinary tags, and the spatial gap between the two different groups seems consistent with the point at which the empirical result of the lower (a) in Fig.~\ref{fig:pdf.empirical} deviates from the theoretical curve. It suggests one hypothesis---a {\it constant rate growth} rather than growth by preferential attachment. With constant rate growth, the selection probability of tags is assumed to be constant. In other words, it is independent of the number of their potential ``competitors'' implicitly considered in the proportional tag selection to their popularity. This is simply defined as follows:
\begin{align}
  n_i(t+1)=n_i(t)+\bar{\alpha}c,
\end{align}
where $c$ is a constant growth rate and $\bar{\alpha}$ is approximately one, as before. Then we obtain the solution $n_i(t)=c(t-t_i)+1$, and after some straightforward substitutions, the expected value of $y'$ in the constant rate growth is obtained as follows:
\begin{align}
  y'_c=ct_i\sim c\hat{t},
\end{align}
assuming $\hat{t}\gg\hat{n}$, where $\hat{t}$ is the expected timepoint at which $\hat{n}$ is reached. The black dashed lines shown for eye guide in Fig.~\ref{fig:heatmap.t_vs_y} are $y'=\hat{t}/10^4$ in each figure. They roughly correspond to the time-dependent structure of the rapidly growing group, suggesting that the growth follows the constant rate growth instead of the preferential attachment. This is more salient in the early stages of growth; meanwhile, it disappears as $\hat{n}$ increases.

\begin{figure*}[t]
\centering
\includegraphics[width=1\linewidth]{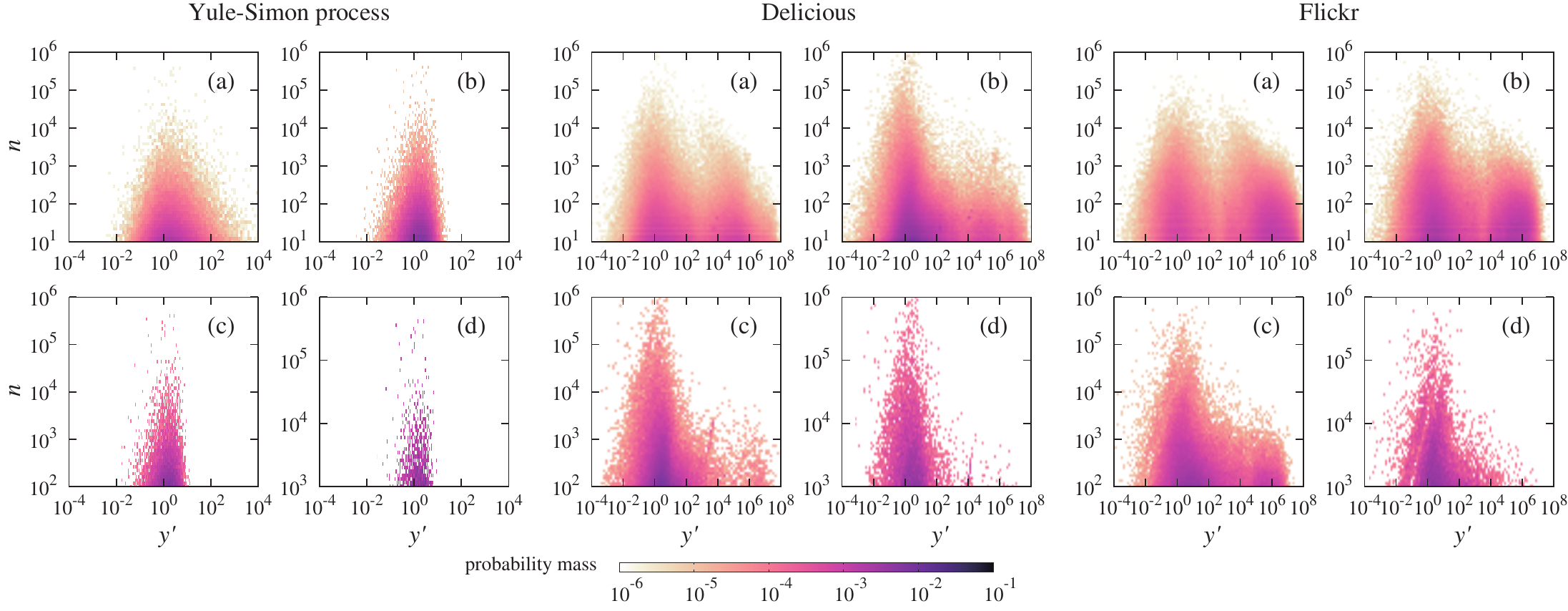}
\caption{Relationship between $y'$ and the final cumulative number of occurrences $n$ of qualified tags. (a) to (d) correspond to the cases of $\hat{n}=2,10,10^2,10^3$, respectively. The color hue indicates the probability mass of tags counted in the heatmap cell, normalized by the total number of qualified tags. The cell size is log-binned with 100 bins both vertically and horizontally between those max and min values.}
\label{fig:heatmap.y_vs_n}
\end{figure*}

This brings us to the question of whether a rapidly growing group or a constant rate growth group accounts for the tags with larger popularity. Figure \ref{fig:heatmap.y_vs_n} shows the relationship between $y'$ and $n$, the final cumulative number of occurrences of tags. The empirical result exhibits two peaks around $y'=1$ and $10^4\sim 10^8$, which is particularly clear in the cases of smaller $\hat{n}$, by which we will readily understand that those two peaks correspond to the two groups that we observed before. That is, the left and right peaks are the tags following, respectively, the preferential attachment and the constant rate growth. However, looking at the left peak in detail, its edges are relatively wider than those of the Yule--Simon process; hence, we suppose that it has not definitely followed the pure preferential attachment. The mechanism of this large deviation around $y'=1$ will be discussed in the next section. In any case, there exists no simple positive correlation between $y'$ and $n$; rather, in the long run, the tags with $y'$ closer to one---the mean-field solution of the preferential attachment---are located at the top of the left peak, suggesting that such tags are dominant in a popularity hierarchy. Krapivsky and others showed that {\it lead changes} are rare in popularity-driven growing networks, where leadership is restricted to the earliest nodes.~\cite{Krapivsky2002b} Their result seems similar to our result here.

\section{Discussion}
We have seen that the popularity dynamics of tags apparently follow the Yule--Simon process (or its sub-linear preferential attachment type) when looking at their aggregated statistical behavior. However, when focusing on the growing dynamics of individual tags, we found there are a considerable amount of tags that grow extremely rapidly or more slowly than the Yule--Simon process predicts. Further, the growth of some of these tags was reasonably explained by the constant rate growth hypothesis in their early stage; meanwhile, some other tags still exhibit a large deviation even in the long-term growth. Summarizing the point, we have two questions: What is the origin of the constant rate growth and the large deviation around the mean-field solution, and why does their effect not distort the selection statistics shown in Fig.~\ref{fig:heaps_and_zipf_law} (c)?

In terms of the first question, we consider that it can be attributed to the cognitive nature of attention. We human beings do not have the ability to hold and handle a whole vocabulary or a whole history of activities at once, and our daily attention tends to be focused on the recent events. This kind of cognitive bias has been also referred to as ``attention decay'' in the context of scientific citations~\cite{Parolo2015}, where the number of citations to scientific papers is bound to decay rapidly a few years after its publication. At the same time, our social activities often exhibit bursty behavior~\cite{Barabasi2005}, where successive and intermittent patterns of occurrence demonstrate a non-Poisson nature. If such bursty and bounded use of words is assumed, it could appear as significant use of the word right after its creation. We consider this to be one explanation for the constant rate growth in the early stage of tag growth. Furthermore, this bounded attention will also boost the use of words that remain within our focus of attention, resulting in a large deviation of its growth in the long run.

In terms of the second question, the attention decay is intuitively supposed to change the selection dynamics drastically. For example, Cattuto and others studied the memory effect on preferential attachment~\cite{Cattuto2006}, and they showed that their modification to the selection probability using a specific memory kernel causes sub-linearity such as we observed in the Flickr data, resulting in a stretched exponential popularity distribution. However, we still have the example of Delicious, which follows just the linear preferential attachment, and further, we have observed that such memory effects on a whole set of tags never cause the large deviation shown here. So, such this explanation is rejected in our situations.

However, what if the attention decay happened in just a portion of the tags? In other words, if there coexisted two kinds of tags: ``minor'' tags that are forgotten rapidly after their creation and ``authoritative'' tags that do not lose their attractiveness. This provides a more reasonable explanation both for the linear and sub-linear preferential attachment and the large deviation around the mean-field solution. That is, the minor and authoritative tags are considered to account for, respectively, the slow and rapid sides of the deviation around the mean-field solution, since the number of competitors of authoritative tags is smaller than the case in which all words survive. Then, if the proportion of minor and authoritative tags is kept constant throughout\footnote{Or at least for smaller $\hat{n}$ than where the weighted frequency starts declining due to the finite size effect of the system.} the class hierarchy in Fig.~\ref{fig:heaps_and_zipf_law} (c), the selection probability will seem superficially to follow a similar curve to that without such distinction.

\section{Conclusion}
We shed light on a novel aspect of social tagging dynamics---a large deviation in the popularity growth of individual tags, based on both theoretical and empirical approaches. We derived the probability density functions of the deviation scaling factors, which describe the tag-independent nature of the popularity growth including the innate fluctuation of the Yule--Simon process. Using those scaling factors, the actual web services were evaluated and were shown to exhibit a large deviation from the mean-field solution beyond the expectation. This anomaly may be attributed to bursty use of new words after their creation and preferential attachment based on attention inequality. Furthermore, such anomalies do not necessarily contribute to the absolute value of the popularity; rather, the mean-field behavior of the preferential attachment is dominant among the most popular tags. We gave some tentative rationalizations of our new findings, which need further analysis to be verified. We also expect that our methods and results here might provide a basis for advanced analysis of other systems such as social networks, citing systems, web evolution, or more general, complex networks~\cite{Newman2006b} that have been studied using the framework of preferential attachment dynamics.

\section*{Acknowledgements}
We would like to express our appreciation to Hiroki Kojima and Koya Sato, our lab colleagues, for their helpful comments.
This work was financially supported by JSPS KAKENHI Grant Number 16K00418 and MEXT
as Post-K Computer Exploratory Challenges---Construction of models for interaction among multiple socioeconomic phenomena (hp160264).

\appendix
\section*{Appendix}
We derive the probability density distribution of the deviation scaling factors in detail. The derivation of the size deviation scaling factor is straightforward from \cite{Hashimoto2016}. The time (timepoint and elapsed-time) deviation scaling factor is the extended idea in this work from which we may elicit further information about the growth of tags.

\subsection{Size deviation distribution}
\label{sec:px}
We denote the probability of the cumulative number of occurrences of tag $i$ at time $t$ being $n$ as $P[n_i(t)=n]$. With sufficiently small $\alpha$, the probability is transformed as follows:
\begin{align}
  \lim_{\alpha\rightarrow 0}P[n_i(t)=n]&=\frac{t_i}{t}\left(1-\frac{t_i}{t}-\frac{n}{t}+\frac{2}{t}\right)^{n-1}\nonumber\\
  &\sim\frac{t_i}{t}\left(1-\frac{t_i}{t}\right)^{n-1},
  \label{eq:P_to_be_n_at_t}
\end{align}
assuming $t\gg n$ and $t\gg 1$. Now we measure $n_i(t)$ for every tag at a specific timepoint, which varies with the birth time of each tag. Considering the expected timepoint to reach $\hat{n}$, the observation timepoint is set to $\hat{n}t_i$ for tag $i$ from Eq.~(\ref{eq:growth_solution}). Introducing a scaling factor of the size deviation, say $x$, we consider the probability that $n_i(t)$ is equal to $x$ times the value of $\hat{n}$ at time $\hat{n}t_i$ as follows:
\begin{align}
  P[n_i(\hat{n}t_i)=x\hat{n}]&\sim\frac{1}{\hat{n}}\left(1-\frac{1}{\hat{n}}\right)^{x\hat{n}-1},
  \label{eq:P_to_be_xn_at_t}
\end{align}
where we replaced $t$ and $n$ in Eq.~(\ref{eq:P_to_be_n_at_t}) with $\hat{n}t_i$ and $x\hat{n}$, respectively. As we are moving from a discrete variant $n$ to a continuous variant $x$, we consider a probability density function for $x$ by assuming the slow rate of change of the probability for $n$, as follows:
\begin{align}
  p(x)&\equiv P[n_i(\hat{n}t_i)=x\hat{n}]\frac{dn}{dx}\nonumber\\
  &\sim\left(1-\frac{1}{\hat{n}}\right)^{x\hat{n}-1},
  \label{eq:PD_to_be_xn_at_t}
\end{align}
where we used $dn/dx=\hat{n}$. This result does not depend on $i$ (i.e., individual tags) and tells us that the probability decays exponentially with the scaling factor in the large limit of $\hat{n}$.

\subsection{Time deviation distribution}
\label{sec:py}
On the other hand, the probability that the cumulative number of occurrences of tag $i$ becomes $n$ right at time $t$ is given as follows:
\begin{align}
  P[n_i(t)\rightarrow n]\equiv P[n_i(t-1)=n-1]\left(\frac{n-1}{t-1}\right).
  \label{eq:P_to_become_n_at_t}
\end{align}
Substituting Eq.~(\ref{eq:P_to_be_n_at_t}) into (\ref{eq:P_to_become_n_at_t}), we obtain
\begin{align}
  P[n_i(t)\rightarrow n]&\sim\frac{t_i}{t-1}\left(1-\frac{t_i}{t-1}\right)^{n-2}\left(\frac{n-1}{t-1}\right)\nonumber\\
  &\sim\left[\frac{t_i(n-1)}{t^2}\right]\left(1-\frac{t_i}{t}\right)^{n-2},
  \label{eq:P_to_become_n_at_t_approx}
\end{align}
assuming $t\gg 1$. Now we fix $n$ to $\hat{n}$, and for every tag, we measure the timepoint at which $n_i(t)$ reaches $\hat{n}$, wherein the expected timepoint should be $\hat{n}t_i$ as was used in the size deviation distribution. Introducing a scaling factor for the timepoint deviation, say $y$, we consider the probability that $n_i(t)$ reaches $\hat{n}$ at $y$-times faster than expected as follow:
\begin{align}
  P\left[n_i\left(\frac{\hat{n}t_i}{y}\right)\rightarrow \hat{n}\right]\sim \frac{y(\hat{n}-1)}{\hat{n}t}\left(1-\frac{y}{\hat{n}}\right)^{\hat{n}-2}.
  \end{align}
Note that the prerequisite $t_i\le t$ bounds the upper limit of $y$ to $\hat{n}$. Again, we move to the new measure from discrete time to the continuous scaling factor of time; hence, we consider a probability density function for $y$ by assuming the slow rate of change of the probability for $t$, as follows:
\begin{align}
  p(y)&\equiv P\left[n_i\left(\frac{\hat{n}t_i}{y}\right)\rightarrow \hat{n}\right]\frac{dt}{dy}\nonumber\\
  &\sim \frac{\hat{n}-1}{\hat{n}}\left(1-\frac{y}{\hat{n}}\right)^{\hat{n}-2},
  \label{eq:PD_to_be_yt}
\end{align}
where $dt/dy=-\hat{n}t_iy^{-2}$. This result also tells us that the probability density of $y$ is independent of individual tags.

Furthermore, we consider another aspect of the time deviation distribution---the elapsed time (not the timepoint). The new scaling factor, say $y'$, of the elapsed-time deviation is defined as follows:
\begin{align}
  y'\equiv \frac{\hat{n}t_i-t_i}{\hat{n}t_i/y-t_i}=\frac{y(\hat{n}-1)}{\hat{n}-y}.
  \label{eq:another_time_scale_factor}
\end{align}
A schematic diagram depicting the timepoint and elapsed-time deviation from the expected growth is shown in Fig.~\ref{fig:diagram} of the main text. The difference between $y$ and $y'$ is only a matter of view, and they have a single-valued relation. For example, both conditions $y=1$ and $y'=1$ yield Eq.~(\ref{eq:growth_solution})---absence of any deviation. However, $y'$ has no upper limit; meanwhile, $y$ is bounded to $\hat{n}$ as mentioned above. This feature of $y'$ is considered an advantage when looking into large deviations in detail. Following the same derivation as from Eq.~(\ref{eq:P_to_become_n_at_t}) to (\ref{eq:PD_to_be_yt}), we obtain the probability density distribution for $y'$ as follows:
\begin{align}
  p(y')&\sim \left(\frac{\hat{n}-1}{y'+\hat{n}-1}\right)^{\hat{n}}.
  \label{eq:PD_to_be_y2t}
\end{align}
Naturally, Eq.~(\ref{eq:PD_to_be_y2t}) can also be derived directly from Eq.~(\ref{eq:PD_to_be_yt}) by multiplying $dy/dy'$ to it.

Lastly, we should note that when measuring the time deviation distribution in data, the finite size effect of data must be taken into account. Thus, $y$ and $y'$ can only be measured for tags that reach a given $\hat{n}$. That is, the larger $y$ or $y'$ becomes, the more such tags will fall within the scope of our observations. This observation bias requires a correction term $y$ to be multiplied into Eqs.~(\ref{eq:PD_to_be_yt}) and (\ref{eq:PD_to_be_y2t}) as follows:
\begin{align}
  q(y)&\equiv yp(y)\nonumber\\
  &\sim y\left(\frac{\hat{n}-1}{\hat{n}}\right)\left(1-\frac{y}{\hat{n}}\right)^{\hat{n}-2}
  \label{eq:PD_to_be_yt_with_correction}
\end{align}
for the timepoint deviation distribution, and
\begin{align}
  q(y')&\equiv yp(y')\nonumber\\
  &\sim y\left(\frac{\hat{n}-1}{y'+\hat{n}-1}\right)^{\hat{n}}\nonumber\\
  &= \frac{\hat{n}y'}{y'+\hat{n}-1}\left(\frac{\hat{n}-1}{y'+\hat{n}-1}\right)^{\hat{n}}
  \label{eq:PD_to_be_y2t_with_correction}
\end{align}
for the elapsed-time deviation distribution. These corrected forms are the effective probability density distributions that we actually observe in the finite data.

\begin{figure*}[ht]
\centering
\includegraphics[width=.8\linewidth]{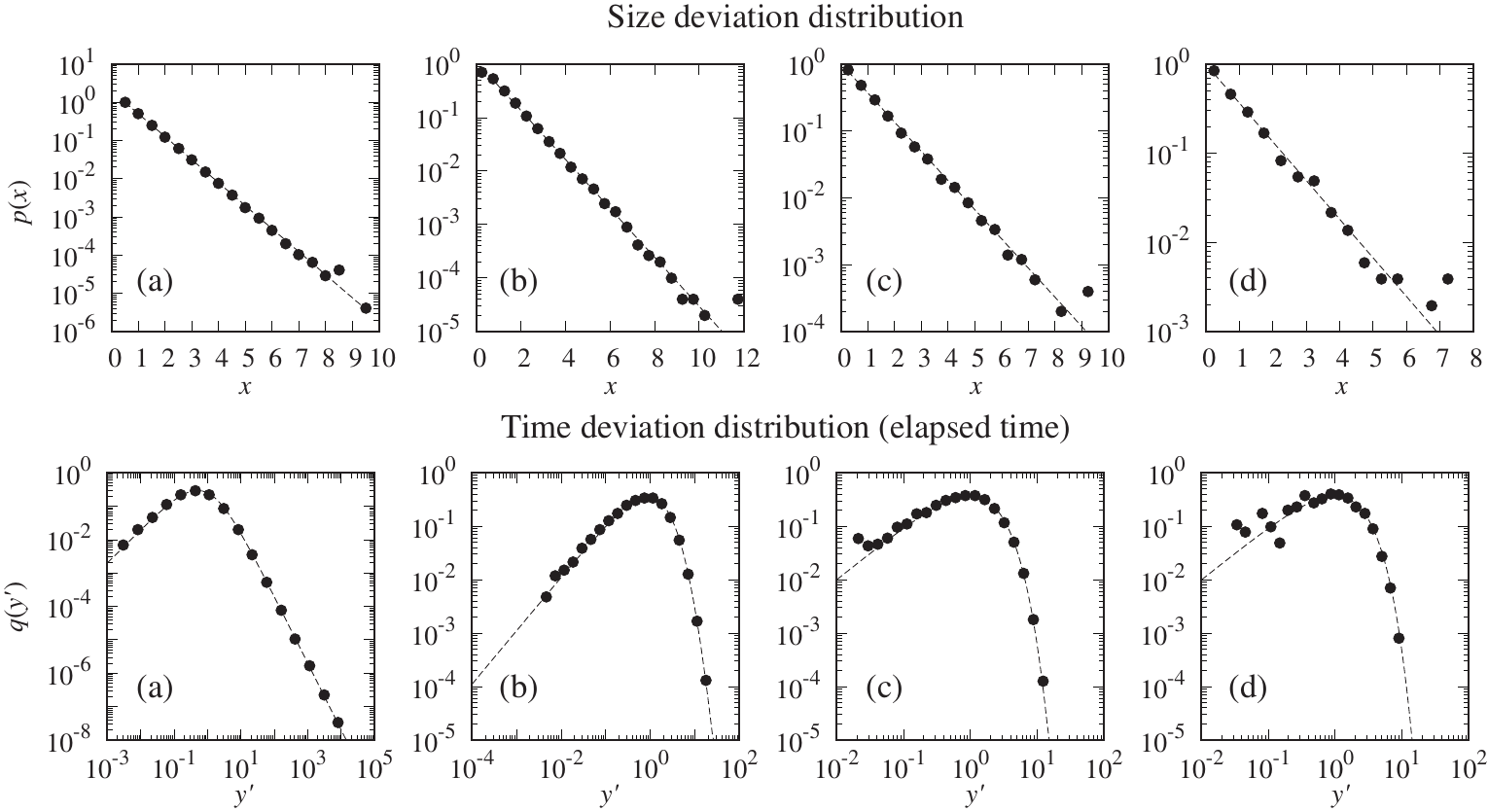}
\caption{Probability density distribution of the size and time deviation scaling factors in the Yule--Simon process. (a) to (d) correspond to the cases of $\hat{n}=2,10,10^2,10^3$, respectively. Black dots and dashed lines show the simulation results and theoretical curves (\ref{eq:PD_to_be_xn_at_t}) for the size deviation distribution and (\ref{eq:PD_to_be_y2t_with_correction}) for the time deviation distribution (elapsed time), respectively. When counting the probability density, linear and logarithmic binning was used, respectively, for $p(x)$ and $q(y')$ with 20 bins between the max and min values.}
\label{fig:pdf.Yule-Simon}
\end{figure*}

\subsection{Numerical validation}
We check the validity of the three probability density functions (\ref{eq:PD_to_be_xn_at_t}), (\ref{eq:PD_to_be_yt_with_correction}), and (\ref{eq:PD_to_be_y2t_with_correction}) via numerical experiments. The experiments are carried out with a long sequence of tags generated by a numerical simulation of the Yule--Simon process; the length of the sequence is $N=10^8$ and the novelty rate $\alpha$ is fixed throughout the simulation to a small value, $\alpha=10^{-2}$. This setup is comparable with the real data we see in the empirical analysis. Once we have built a tag sequence, values of $x$, $y$, and $y'$ are measured for individual qualified tags, varying $\hat{n}$ as $2,10,10^2,10^3$. The qualification of tags is to live $\hat{n}t_i$ time steps since its birth in measuring $x$, or to reach $\hat{n}$ by the end of the simulation in measuring $y$ and $y'$. Actually, the total number of qualified tags was approximately $\alpha N/\hat{n}$ in all measurements of $x$, $y$, and $y'$ with different $\hat{n}$, where $\alpha N$ is the expected vocabulary size. This is a reasonable consequence of measuring $x$ since the birth time of each tag is distributed uniformly over the sequence; meanwhile, it is not so innately obvious in measuring $y$ and $y'$.

Figure.~{\ref{fig:pdf.Yule-Simon}} shows the probability density distribution of the size and time deviation scaling factors obtained by the simulation. As $\hat{n}$ increases, the number of qualified tags decreases and the simulation results become somewhat noisy. However, they exhibit an overall good agreement with the theoretical curves we obtained.

\end{document}